\documentclass[useAMS,usenatbib]{mn2e}
\usepackage{graphicx}
\usepackage{amssymb}
\newcommand{\Teff}{\mbox{$T_{\mathrm{eff}}\,$}}
\newcommand{\Logg}{\mbox{$\log\,g\,$}}
\newcommand{\Mwd}{\mbox{$M_{\mathrm{wd}}$}}
\newcommand{\Rwd}{\mbox{$R_{\mathrm{wd}}$}}
\newcommand{\Bwd}{\mbox{$B_{\mathrm{wd}}$}}
\newcommand{\Msun}{\mbox{$\mathrm{M}_{\odot}$}}

\newcommand{\MiPG}{\mbox{$M_i^\mathrm{PG}$}}
\newcommand{\MiSDSS}{\mbox{$M_i^\mathrm{SDSS}$}}
\newcounter{tref}
\newcommand{\tbr}{$^{\arabic{tref}}$\stepcounter{tref}}
\newcommand{\tbc}{\arabic{tref}\stepcounter{tref}}

\title{PG\,1258+593 and its common proper motion magnetic white dwarf counterpart}

\author[J.Girven et al.]{J. Girven$^1$, B. T. G\"ansicke$^1$,
  B. K\"ulebi$^2$, D. Steeghs$^1$, S. Jordan$^2$, T.R. Marsh$^1$, D. Koester$^3$\\
$^{1}$ Department of Physics, University of Warwick, Coventry CV4 7AL,
UK \\ 
$^{2}$ Astronomisches Rechen-Institut, Zentrum f\"ur Astronomie
der Universit\"at Heidelberg, M\"onchhofstrasse 12-14, D-69120
Heidelberg, Germany\\
$^{3}$ Institut f\"{u}r Theoretische Physik und Astrophysik, University of Kiel, 24098 Kiel, Germany\\
}

\begin{document}

\date{Started 2009}

\pagerange{\pageref{firstpage}--\pageref{lastpage}} \pubyear{2009}

\maketitle

\label{firstpage}

\begin{abstract}
We identify SDSS\,J130033.48+590407.0 as a common proper motion companion to the
well-studied DA white dwarf PG\,1258+593 (GD322). The system lies at a distance
of $68\pm3$\,pc, where the angular separation of $16.1\pm0.1\arcsec$ corresponds
to a minimum binary separation of $1091\pm7$\,AU. SDSS\,J1300+5904 is a cool
($\Teff=6300\pm300$K) magnetic white dwarf ($B\simeq6$\,MG). PG\,1258+593 is a
hydrogen-rich (DA) white dwarf with $\Teff=14790\pm77$\,K and $\log
g=7.87\pm0.02$. Using the white dwarf mass--radius relation implies the masses
of SDSS\,J1300+5904 and PG\,1258+593 are $0.54\pm0.06$\,\Msun\ and
$0.54\pm0.01$\,\Msun, respectively, and therefore a cooling age difference of
$1.67\pm0.05$\,Gyr. Adopting main-sequence life times from stellar models, we
derive an upper limit of 2.2\,\Msun\ for the mass of the progenitor of
PG\,1258+593. A plausible range of initial masses is 1.4--1.8\,\Msun\ for
PG\,1258+593 and 2--3\,\Msun\ for SDSS\,J1300+5904. Our analysis shows that
white dwarf common proper motion binaries can potentially constrain the white
dwarf initial-final mass relation and the formation mechanism for magnetic white
dwarfs. The magnetic field of SDSS\,J1300+5904 is consistent with an Ap
progenitor star. A common envelope origin of the system cannot be excluded, but
requires a triple system as progenitor.
\end{abstract}


\section{Introduction}
\label{s-int}

White dwarfs (WDs) are the end point for the large majority of
stars. It has been shown that a significant number, possibly
$\sim10-15\%$, of all WDs may be magnetic with fields $\ga1$\,MG
\citep{liebertetal03-1, wickramasinghe+ferrario05-1}. The Sloan
Digital Sky Survey (SDSS; \citealt{yorketal00-1}) has been a rich
source for finding new magnetic WDs (MWDs)
(\citealp{gaensickeetal02-5, schmidtetal03-1, vanlandinghametal05-2},
and \citealp{kuelebietal09-1}), bringing the number of known MWDs to
$>200$.  However, the formation mechanism for MWDs is still under
debate, with the two favoured progenitors being either magnetic Ap/Bp
stars \citep{moss89-1} or close binaries that evolved, and potentially
merged, through a common envelope \citep{tout+pringle92-1}.

In the Ap/Bp scenario, the MWDs field is a relic of the large-scale
magnetic fields of their intermediate mass progenitor stars. These in
turn are fossils of the magnetic field in star formation
\citep{moss89-1}. Assuming flux conservation, the surface fields
observed in Ap/Bp stars ($\sim10^2-2\times10^4$\,G) are sufficient to
explain the range of fields found in MWDs. However, population
synthesis suggest that only 40\% of the known MWDs may have descended
from Ap/Bp stars \citep{wickramasinghe+ferrario05-1}.

A clue to a possible link between binary evolution and strongly
magnetic WDs came from the \textit{absence} of detached MWD plus
M-dwarf binaries, i.e.  magnetic pre-CVs \citep{liebertetal05-2},
which could not be explained within the Ap/Bp scenario.  Differential
rotation and convection are predicted to be key to a magnetic dynamo
\citep{tout+pringle92-1}, both of which are prevalent in common
envelope (CE) evolution. \citet{toutetal08-1} recently revisited the
CE scenario for the formation of MWDs, and proposed that if a strong
field is generated during a CE, the two possible outcomes are either a
merger, leading to a single massive, strongly magnetic WD, or a
short-period MWD plus low-mass star binary, that rapidly evolve into a
mass-transferring CV state.

A key for testing which of the hypotheses is correct would be a set of
wide common proper motion (CPM) MWDs. Here, we report the discovery of
one such system.  SDSS\,J130033.48+590407.0 (henceforth
SDSS\,J1300+5904) is a MWD and the CPM companion to the well-studied
hydrogen-rich (DA) WD PG\,1258+593. Following an estimate of the
stellar parameters of SDSS\,J1300+5904, we discuss the evolutionary
state of the CPM pair, and show that in this system, the Ap/Bp
scenario provides a plausible explanation for the origin of the
MWD. We also illustrate how WD CPM pairs may be used to constrain
semi-empirical initial mass-final mass relations (IFMR) at the
low-mass end.

\begin{table*}
\caption{\label{t-obs} Coordinates, proper motions, and PSF magnitudes
  of the two WDs extracted from SDSS DR7.}
\setlength{\tabcolsep}{0.8ex}
\begin{tabular}{lcccccccccc}
\hline
Object & RA & Dec & \multicolumn{2}{c}{p.m. [mas\,yr$^{-1}$]} & $u$ & $g$ & $r$ & $i$ & $z$ \\
& (2000) & (2000) & RA & Dec \\\hline
PG\,1258+593 & 13 00 35.20 & +59 04 15.6 & $42.4\pm2.6$ & $75.0\pm2.6$ &
$15.54\pm0.01$ & $15.20\pm0.04$ & $15.52\pm0.02$ & $15.76\pm0.04$ & $16.04\pm0.02$ & \\
SDSS\,J1300+5904 & 13 00 33.46 & +59 04 06.9 & $41.8\pm3.0$ & $73.9\pm3.0$ &
$19.08\pm0.03$ & $18.23\pm0.04$ & $17.93\pm0.02$ & $17.80\pm0.04$ & $17.79\pm0.03$ \\
\hline
\end{tabular}
\end{table*}

\section{Observations}
\label{s-obs}

\begin{figure}
\includegraphics[width=\columnwidth]{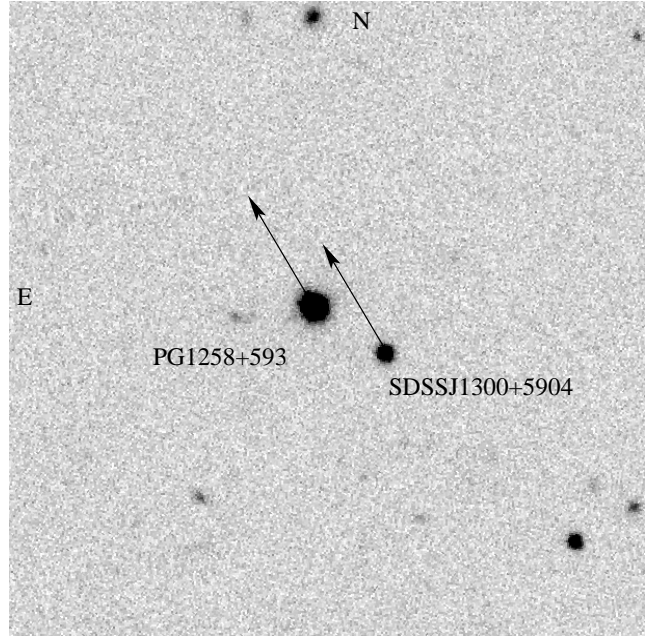}
\caption{\label{f-1300image} $2\arcmin\times2\arcmin$ $r$ band SDSS image
  of the WD CPM pair. Proper motions are indicated over 300 years. An SDSS fibre spectrum was obtained for the fainter 
  component of the binary.}
\end{figure}

One of the novelties within SDSS Data Release 7~(DR7)
\citep{abazajian09-1} are improvements in both the astrometric
calibration, carried out against the UCAC2 catalogue
\citep{zachariasetal04-1}, as well as an updated table of proper
motions computed from the combined SDSS and USNO-B
\citep{monetetal03-1} positions. Using this proper motion table, we
carried out a search for $3\sigma$ proper motion companions to WDs using the CasJobs SQL interface to SDSS~DR7
\citep{li+thakar08-1}. One of the objects returned by our query was
the faint blueish SDSS\,J1300+5904, which turned out to be a CPM
companion to PG\,1258+593\footnote{Initially identified as WD
  candidate GD\,322 by \citet{giclasetal67-1}}. The angular separation
between the two objects is $16.1\pm0.1\arcsec$
(Fig.\,\ref{f-1300image}) and, at a distance of $68\pm3$\,pc (see
Sect.\,~\ref{s-wdparam}), the minimum binary separation is
$1091\pm7$\,AU.  Coordinates, proper motions, and $ugriz$ point-spread
function (PSF) magnitudes of both WDs are given in
Table\,\ref{t-obs}.

SDSS\,J1300+5904 had already been noted as a CPM companion by \citet{farihietal05-1}, however it was
classified as a WD with a featurless (DC) spectrum, based on a
relatively poor spectrum. Inspecting the SDSS fibre spectrum however unambiguously
identifies it as a magnetic (DAH) WD given the clear
detection of a Zeeman-triplet in H$\alpha$
(Fig.\,\ref{f-1300fit}). SDSS\,J1300+5904 was targeted for
SDSS spectroscopy as a WD candidate; no SDSS spectrum was
obtained for PG\,1258+593.

We observed PG\,1258+593 on February 13, 1997, using the
Intermediate Dispersion Spectrograph (IDS) on the Isaac Newton Telescope
(INT). Two spectra of 20\,min exposure time each were obtained with
the R632V grating and a 1.5$\arcsec$ slit, covering the wavelength
range 3680--5300\,\AA\ at a spectral resolution of $\sim2.3$\,\AA. The
data were reduced and calibrated as described by \citet{moranetal99-1}, and the
normalised line profiles are shown in Fig.\,\ref{f-pgfit}.

\section{White dwarf parameters}
\label{s-wdparam}

\begin{figure}
\includegraphics[width=\columnwidth]{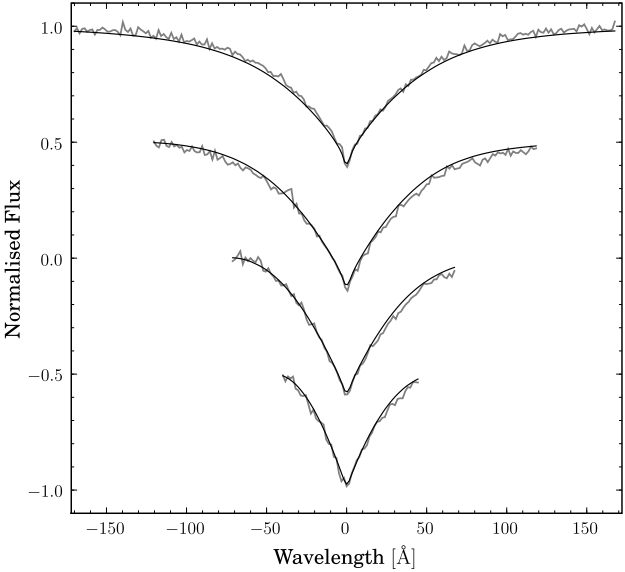}
\caption{\label{f-pgfit} Normalised INT/IDS H$\beta$--H$\epsilon$ (top
  to bottom) line profiles of PG\,1258+593 (gray line) and the
  best-fit model (black line) for $\Teff=14790$\,K and $\log
  g=7.87\pm0.02$. }
\end{figure}

\begin{figure}
\includegraphics[width=\columnwidth]{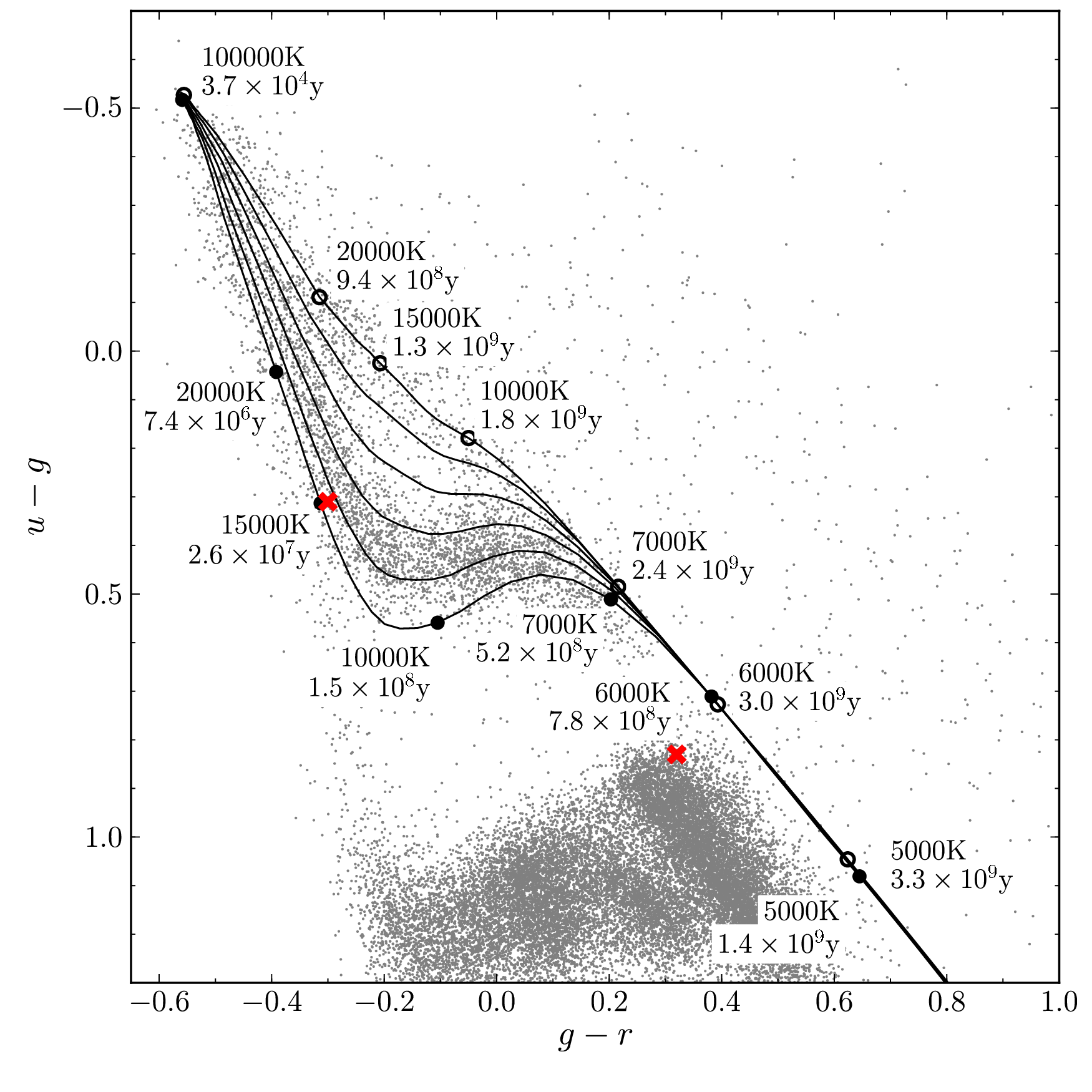}
\caption{\label{f-cc} SDSS $u-g$ vs $g-r$ colour-colour diagram
  showing PG\,1258+593 (left) and its magnetic CPM companion
  SDSS\,J1300+5904 (right) as red crosses. Theoretical DA cooling
  tracks shown as black lines for (from left to right) $\log g=7-9.5$
  in steps of 0.5. The black open and black filled circles track
  $\log g=9.5$ and $\log g=7.0$ respectively, with corresponding cooling
  times.}
\end{figure}

We analysed the INT/IDS spectrum of PG\,1258+593 using DA model
spectra from \citet{koesteretal05-1} and the fitting routine described
by \citet{rebassa-mansergasetal07-1}. The best fit is achieved for
$\Teff=14790\pm77$\,K and $\Logg=7.87\pm0.02$.  Adopting these
atmospheric parameters, we use an updated version of
\citeauthor{bergeronetal95-2}'s \citeyearpar{bergeronetal95-2}
tables\footnote{http://www.astro.umontreal.ca/$\sim$bergeron/CoolingModels/}
to calculate the corresponding WD mass, $0.54\pm0.01\,\Msun$, radius,
$(9.85\pm0.10)\times10^8$\,cm, and a cooling age of
$(1.8\pm0.07)\times10^8$\,yr.  Finally, we calculate
$M_g=11.03\pm0.1$, corresponding to a distance of $68\pm3$\,pc
(Table\,\ref{t-wdprop}). The best fit is shown in
Fig.\,\ref{f-pgfit}. The $u-g$ vs $g-r$ colours of PG\,1258+593 are
broadly consistent with the results of the spectroscopic analysis
(Fig.\,\ref{f-cc}). Our atmospheric and stellar parameters for
PG\,1258+593 are in excellent agreement with those published by
\citet{liebertetal05-1} as part of their systematic analysis of the DA
WDs from the Palomar Green Survey. They quote $\Teff=14480\pm229$\,K,
$\Logg=7.87\pm0.05$, and $\Mwd=0.54\pm0.02\,\Msun$.

Given the magnetic nature of SDSS\,J1300+5904, establishing its
atmospheric parameters is not straight-forward, and we analysed the
SDSS photometry and spectroscopy with both non-magnetic and magnetic
model spectra. 

\begin{figure*}
\includegraphics[width=0.9\textwidth]{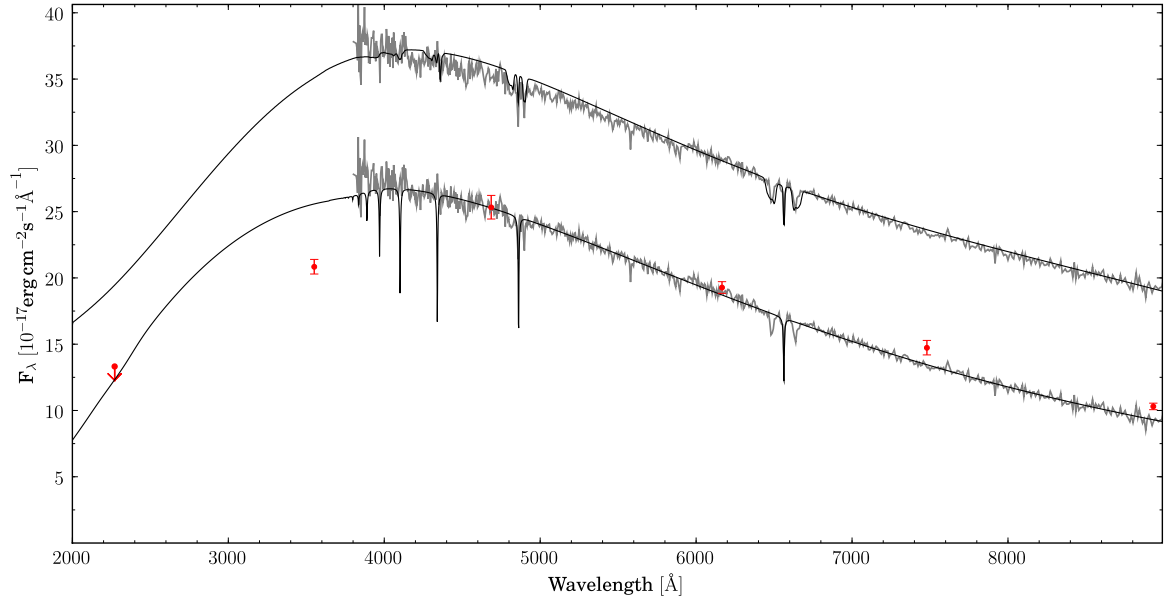}
\caption{\label{f-1300fit} The SDSS spectrum of SDSS\,J1300+5904 along
  with non-magnetic (bottom curves, $\Teff=6500\,K$, $\Logg=7.93$)
  and magnetic (top curves, $\Teff=6000$\,K, $\log g=7.93$) WD
  models. The top curves are offset by 10 flux units. The magnetic WD
  model is calculated for a centered dipole with polar strength of
  6\,MG at an inclination against the line-of-sight of $45$
  degrees. Shown in red are the fluxes corresponding to the SDSS
  $ugriz$ PSF magnitudes of SDSS\,J1300+5905. The left-most red point
  indicates the upper limit on the near-ultraviolet flux of
  SDSS\,J1300+5905 implied by the non-detection with GALEX.}
\end{figure*}

Figure\,\ref{f-cc} shows the location of SDSS\,J1300+5904 in the $u-g$
vs $g-r$ colour plane, which, while being somewhat displaced from the
cooling tracks of non-magnetic DA WDs, clearly suggest a low
temperature. As a first step, we performed a least $\chi^2$ fit to the
SDSS $ugriz$ magnitudes using model DA colours from
\citet{koesteretal05-1}, which results in a temperature estimate of
6000\,K. As expected from the morphology of the DA cooling tracks
(Fig.\,\ref{f-cc}), at such low temperatures, the colours provide very
little information on \Logg. Taking the SDSS observations at face
value, the $u$-band flux from the SDSS imaging appears somewhat too
low compared to the flux level of the SDSS fibre spectrum
(Fig.\,\ref{f-1300fit}). Such offsets between the SDSS spectroscopy
and photometry are found in a number of objects, and are in most cases
strongest in the $u$-band. The discrepancy seen in SDSS\,J1300+5904 is
consistent with the location of SDSS\,J1300+5904 in the $u-g$ vs $g-r$
colour plane (Fig.\,\ref{f-cc}), where the $u-g$ colour of the object
is too red with respect to the DA cooling tracks. A reduced $u$ band
flux with respect to the extrapolation of the spectrum could otherwise
be caused by a large Balmer jump, contrary to that expected for a
$\sim6000$\,K DA WD. In reality, magnetic splitting probably leads to
a shallower Balmer jump than in the non-magnetic case. Photometric
variability due to rotation, such as observed e.g. in the magnetic WD
GD\,356 \citep{brinkworthetal04-1}, also cannot be ruled out, and
further study is warranted.

In a second step, we made use of the known distance, $d=68\pm3$\,pc.  For a given
choice of \Teff, we vary $\log g$ to match the observed SDSS magnitudes, and the
best-fit $\log g$ then provides \Mwd\ and \Rwd\ by adopting a WD mass-radius
relation \citep{wood95-1, fontaineetal01-1}. In the light of the flux
discrepancy in the $u$-band discussed above, we restrict the fit to the $griz$
magnitudes, and find $\Teff=6300\pm300$\,K, $\log g=7.93\pm0.13$, corresponding
to a WD mass $\Mwd=0.54\pm0.06\,\Msun$, radius
$\Rwd=(9.33\pm0.64)\times10^8$\,cm,
and cooling age of $1.7{+0.4\atop-0.2}\times10^9$\,yr.

PG\,1258+593 is detected by GALEX \citep{martinetal05-1} at
$m_\mathrm{fuv}=15.30\pm0.02$ and
$m_\mathrm{nuv}=15.33\pm0.01$. Adopting $\Teff$ and $\Logg$ from
Table\,\ref{t-wdprop}, and $d=68\pm3$\,pc from above, we folded a DA model
spectrum from \citep{koesteretal05-1} through the GALEX far and
near-ultraviolet response curves, obtaining $m_\mathrm{fuv}=15.37$ and
$m_\mathrm{nuv}=15.30$, in excellent agreement with the GALEX
measurement when taking into account the low, but non-zero amount of
reddening along the line of sight and the systematic uncertainties in
the GALEX calibration. In contrast, SDSS\,J1300+5904 is not detected
by GALEX. For an assumed distance of 68\,pc, the limiting magnitude of
GALEX, $m_\mathrm{nuv}=20.5$, implies upper limits for \Teff between
$6350$ and $6650\,\mathrm{K}$ for a mass range from $0.47$ to
$0.63\,\Msun$, which is consistent with the results we obtained from
fitting the $griz$ magnitudes for the same distance.

We also fitted non-magnetic DA model spectra to the observed spectrum
of SDSS\,J1300+5904. This results in $\Teff\simeq6500$\,K, which
corroborates the low temperature suggested by the photometry, but can
obviously not properly account for the observed Balmer line profiles
(Fig.\,\ref{f-1300fit}).

Finally, fixing the distance to $68\pm3$\,pc, and $\log g=7.93$, we analysed
the spectrum of SDSS\,J1300+5904 with magnetic WD models, using a
simplified version of the code explained in \citet{euchneretal02-1}
and following the procedure outlined in \citet{kuelebietal09-1} to fit
for a centred magnetic dipole.  Due to the lack of a consistent theory
that describes Stark broadening in the presence of magnetic fields in
this regime \citep[e.g.][]{jordan92-1} the computed line profiles are
subject to systematic uncertainties. Hence discrepancies between the
apparent strengths of the Balmer lines and the slope of the continuum
are observed \citep[see][]{achilleosetal91-1}. We have used the
approach of \citet{gaensickeetal02-5} and used two different methods
to asses the effective temperature: Fitting only the Balmer lines
(6000\,K) and fitting the continuum slope (6800\,K). In the case where
only Balmer lines are fitted the slope of the model spectrum is
normalized with respect to the observed one.  The Zeeman splitting
observed in H$\alpha$ implies a magnetic field strength of
$\simeq6$\,MG and suggests an intermediate inclination between the
line-of-sight and the magnetic axis. Figure\,\ref{f-1300fit} shows a
magnetic model spectrum for a centered dipole with polar strength of 6
MG with an inclination of $\sim45$ degrees as an example of a
satisfying fit.

\begin{table}
\caption{\label{t-wdprop} Atmospheric and stellar parameters for
  PG\,1258+593 and SDSS\,J1300+5904.}
\begin{tabular}{lcc}
\hline 
     & PG\,1258+593 & SDSS\,J1300+5904 \\
\hline
\Teff [K]     & $14790\pm77$              & $6300\pm300$ \\
\Logg         & $7.87\pm0.02$             & $7.93\pm0.11$ \\
\Mwd [\Msun]  & $0.54\pm0.01$             & $0.54\pm0.06$ \\
\Rwd [$10^8$\,cm] & $9.85\pm0.10$         & $9.33\pm0.64$ \\
Cooling age [$10^8$\,yr] & $1.8\pm0.07$   & $18.5\pm0.5$ \\
\Bwd [MG]     & $\le0.3$ & $\simeq6$ \\
\hline
\end{tabular}
\end{table}

\section{Stellar evolution of the CPM pair}
Within the errors, both WDs in this CPM pair have equal masses,
similar or slightly below the mean mass of DA WDs, $0.593\pm0.016\Msun$
\citep[e.g.][]{koesteretal79-1, finleyetal97-1, liebertetal05-1,
  kepleretal07-1}, but their different effective temperatures result
in an age difference of $1.67\pm0.05$\,Gyr, implying that their
progenitor stars had rather different main-sequence life times.

It is long known that stars undergo different amounts of mass loss
depending on their initial mass, and \citet{weidemann77-1} pioneered
the investigation of the initial-final mass relation (IFMR) for
WDs. The bulk of recent observational work constraining the IFMR has
been carried out using WDs in open clusters spanning a range
of ages \citep[e.g.][]{ferrarioetal05-1, kaliraietal05-1,
  dobbieetal06-1, catalanetal08-1, catalanetal08-2, kaliraietal08-1,
  rubinetal08-1, salarisetal09-1, casewelletal09-1, dobbieetal09-1,
  williamsetal09-1}. These studies exploit the fact that the age of
the cluster population can be determined from the main-sequence
turn-off. The measured WD cooling age can then be used to calculate
the lifetime of the WD progenitor and thus its initial mass can be
estimated.

Clusters, however, are still relatively young, and therefore the low
mass stars have not evolved into WDs yet. This means the low mass end
of the IFMR, below $\sim2$\,\Msun, is very poorly constrained, and the
progenitors of both PG\,1258+593 and SDSS\,J1300+5904 most likely had
initial masses in this range.  The question also remains as to whether
the IFMR is indeed a one-valued relation, or whether there is a
spread.

Wide WD binaries that did not interact during their evolution can in
principle provide additional semi-empirical constraints on the
IFMR. The cooling ages in such binaries can be determined from
standard WD evolution models. The strongest constraints can be
expected to come from binaries containing two WDs with unequal
properties, such that the cooling ages differ significantly. This
method has the advantage that some of the WDs will be of low mass and
thus constrain the low mass end of the IFMR. To our knowledge, such an
approach to the IFMR was attempted only twice.
\citet{greensteinetal83-1} analysed the Sanduleak-Pesch WD binary
(WD\,1704+481), but their results were invalidated by the discovery
that one of the two WDs is itself an unresolved close WD binary that
underwent a common-envelope evolution
\citep{maxtedetal00-1}. \citet{finley+koester97-1} modelled both
components of PG\,0922+162, a CPM binary containing two relatively
massive WDs, and their results are consistent with the IFMR obtained
from open clusters.

\begin{figure}
\includegraphics[width=\columnwidth]{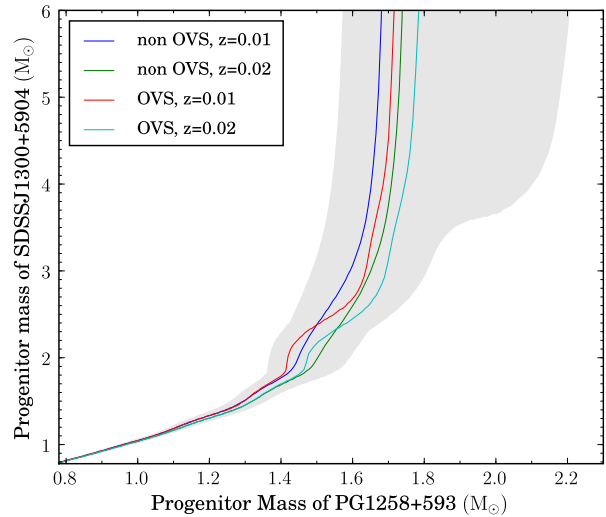}
\caption{\label{f-pmr} The mass of the progenitors star of SDSS\,J1300+5905 as a
function of the mass of the progenitors star of PG\,1258+593, given the
difference between cooling ages indicate the difference between progenitor
lifetimes. Calculations are made for over-shoot (OVS) and non over-shoot models,
as well as solar ($z=0.02$) and half solar ($z=0.01$) metalicities. Allowing for
$1\sigma$ errors on cooling times leads to a spread into the grey shaded region.}
\end{figure}

Here we make use of the age difference between PG\,1258+593 and
SDSS\,J1300+5904 to provide a semi-emirical upper limit on the
progenitor mass of PG\,1258+593. We adopt the main-sequence life times
as a function of initial mass from the stellar evolution models of
\citet{polsetal98-1}. For any given choice of the progenitor mass of
PG\,1258+539 (\MiPG), the age difference of $1.67\pm0.05$\,Gyr then
implies a progenitor mass for SDSS\,J1300+5904
(\MiSDSS). Figure\,~\ref{f-pmr} illustrates the relation between
\MiPG\ and \MiSDSS\ for solar and half-solar metalicity models with
and without overshooting. As the main-sequence life time is a very
strong function of the initial mass, \MiSDSS\ levels off very steeply
for $1.4\la\MiPG\la1.8$, and in a most conservative interpretation,
$\MiPG<2.2$\,\Msun. Being more adventurous, one may choose to adopt
the most recent IFMR cluster relations
\citep[e.g.][]{casewelletal09-1, salarisetal09-1} to turn the mass of
SDSS\,J1300+5904 into a conservative upper limit of
$\MiSDSS<3$\,\Msun, and therefore $\MiPG<1.8$\,\Msun.

Admittedly, for a single WD binary this proves to be merely consistent with the
current IFMR rather than improving it. It may however be in favour of a spread
in the IFMR since we find two WDs with similar masses, yet very different ages
and thus implying different progenitor masses. If we take current IFMRs,
progenitor masses in the range $1-1.4\Msun$ would be expected. The mass errors
in Table\,\ref{t-wdprop} represent only the statistical uncertainty in fitting
the observed Balmer lines with model spectra. Systematic uncertainties in the
models and/or fitting procedure are difficult to assess, but are likely to
outweigh the statistical errors. We estimate that the largest current
mass difference consistent with the observational data is $\simeq0.1$\,\Msun,
which would move, following the procedure outlined above, both progenitor stars
into a range $\simeq1-1.4$\,\Msun.

Our study of PG\,1258+593 and SDSS\,J1300+5904 outlines the potential
of using WD CPM binaries for constraining the IFMR, bearing in mind
that SDSS contains at least a few dozen of such binaries. However, to
fully exploit this method, i.e. to reduce the spread seen in the
relation shown in Fig\,\ref{f-pmr}, high-quality follow-up
spectroscopy plus broad-band photometry are necessary to deliver
accurate \Teff\ and \Logg\ measurements.

A final caveat for the WD CPM binary presented in this study is that
the magnetic field may have affected the IFMR for SDSS\,J1300+5904,
however, there is currently no evidence for such an effect
\citep{wickramasinghe+ferrario05-1, catalanetal08-1}.

\begin{table*}
\caption{\label{t-DDs} Known spatially resolved double degenerate
  systems with one magnetic component.}  \setcounter{tref}{1}
\begin{tabular}{lccccccccl}
\hline\hline
Name & \multicolumn{3}{c}{MWD} & \multicolumn{3}{c}{Companion} & \multicolumn{2}{c}{separation} & Ref.  \\
& \Teff\,[K] & \Mwd [\Msun] & $B$\,[MG] & Type & \Teff\,[K] & \Mwd\,[\Msun] & $[\arcsec]$ & AU & \\ \hline
LB 11146      & $16000$     & $0.9$         & 670 & DA & 14500 & $0.91\pm0.07$ & 0.015 & $\sim0.6$ & \tbc,\tbc,\tbc,\tbc \\
RE J0317-853  & $33800$     & $1.32$        & 340 & DA & 16000 &
$\sim0.93$    & 6.7   & $\sim200$ & \tbc,\tbc,\tbc,\tbc \\
SDSS\,J130033.48+590407.0 & $6300\pm300$ & $0.54\pm0.06$ & $\simeq6$ & DA & $14790\pm77$ & $0.54\pm0.01$ & $16.1\pm0.1$ & $\ga1091$ & \tbc\\ 
\hline
\end{tabular}
\begin{minipage}{\textwidth}
\footnotesize
\setcounter{tref}{1}
\tbr~\citet{liebertetal93-1};
\tbr~\citet{glennetal94-1};
\tbr~\citet{schmidtetal98-1};
\tbr~\citet{nelanetal07-1};
\tbr~\citet{barstowetal95-2};
\tbr~\citet{ferrarioetal97-1};
\tbr~\citet{burleighetal99-1};
\tbr~\citet{vennesetal03-1};
\tbr~this paper.
\end{minipage}
\end{table*}

\section{SDSS\,J1300+5905 and the origin of magnetic white dwarfs} The origin of
highly magnetic ($\ga1$\,MG) WDs is an unsettled issue. Early estimates
of the fraction of MWDs hinted at a value of $\sim4\%$
\citep[e.g.][]{schmidt+smith95-1}, and led to the conclusion that their masses
were on average higher than those of non-magnetic WDs \citep{liebertetal88-1}. 
The space density of MWDs and their high masses were taken as being suggestive
for MWDs descending from chemical peculiar Ap/Bp stars, with the strong fields
of the MWDs explained by magnetic flux conservation \citep[e.g.][]{angel81-1,
toutetal04-1}. However, more recent work suggests that the fraction of MWD may
actually be as high as $10-15\%$ \citep{liebertetal03-1,
wickramasinghe+ferrario05-1}, casting doubt as to whether the space density of
Ap/Bp stars is sufficient for producing all MWDs \citep{kawka+vennes04-1,
wickramasinghe+ferrario05-1}.

\citet{liebertetal05-2} spotted another oddity about MWDs, namely that not a
single MWD has been found in any of the $>1600$ known (wide and close) WD plus
M-dwarf binaries \citep{silvestrietal07-1, helleretal09-1,
rebassa-mansergasetal09-1}~--~contrasting the large frequency of
\textit{interacting} MWD plus M-dwarf binaries, i.e. magnetic cataclysmic
variables, which make up 25\% of all known CVs
\citep{wickramasinghe+ferrario00-1}. This motivated \citet{toutetal08-1} to
outline a very different scenario for the origin of MWDs, in which dynamos
during the common envelope evolution of close binaries generate strong magnetic
fields in the core of the WD progenitor. During the common envelope of a WD plus
M-dwarf binary, the separation shrinks, leading primarily to two different
possible outcomes. If the two stars avoid merging, they leave the common
envelope as a short-period binary that will relatively rapidly start mass
transfer as a magnetic cataclysmic variable. In fact, a number of such systems,
magnetic pre-cataclysmic variables, are known \citep{reimersetal99-1,
reimers+hagen00-1, szkodyetal03-3, schmidtetal07-1, schwopeetal09-1}.
Alternatively, the two stars may coalesce, forming a single MWD, which will
typically be more massive than non-magnetic field WDs.

Given the larger number of observational constraints that are
available for MWDs in binaries, these systems hold a strong potential
in improving our understanding of the origin of MWDs. Until now, only
two spatially resolved WD binaries containing one MWD were known,
RE\,J0317--853 and LB\,11146, see Table\,\ref{t-DDs}. RE\,J0317--853
is hot, massive WD, rotating with a spin period of 725\,sec, and has a
very large magnetic field \citep{barstowetal95-2,
  burleighetal99-1}. Its DA companion LB\,9802 is cooler and of lower
mass\footnote{The physical association of RE\,J0317--853 is so far
  purely based on the small separation on the sky, no proper motions
  are available for the two stars. However, a preliminary analysis of
  HST/FGS data does appear compatible with the system being a CPM pair
  (Jordan et al. in prep.).} than RE\,J0317--853, a paradox in terms
of stellar evolution which is resolved if the magnetic component is
assumed to be a relatively recent merger \citep{ferrarioetal97-1}.
LB\,11146 is a relatively close binary, that might have undergone a
common envelope evolution \citep{nelanetal07-1}. Hence, the properties
of both RE\,J0317--853 and LB\,11146 are consistent with a binary
origin of the magnetic field. In addition to these two spatially
resolved WD plus MWD binaries, about half a dozen unresolved
spectroscopic WD binaries containing an MWD are known \citep[][and
  references therein]{kawkaetal07-1}, and hence it is not known if
they underwent binary interaction or not.

PG\,1258+593/SDSS\,J1300+5904 differ from the two previously known
spatially resolved binaries in that the two WDs appear to have evolved
without interacting, and their properties agree with standard stellar
evolution theory.  Taken the observational facts at face value, it
seems entirely plausible that the strong magnetic field of
SDSS\,J1300+5904 is related to the Ap phenomenon and not related to a
common envelope evolution, unless it is itself an unresolved binary,
or unrecognised merger. The first option appears contrived, but not
impossible (see the case of WD\,1704+481; \citealt{maxtedetal00-1}).
The SDSS spectrum shows no evidence for an additional binary
companion, unless it is a featureless DC WD, similar to that in G62-46
\citep{bergeronetal93-1}, or a very late type dwarf. Using the star
spectral templates of \citet{araujo-betancoretal05-1}, a hypothetical
unresolved late type companion to SDSS\,J1300+5904 has to be of
spectral type L5 or later to go unnoticed. WD plus brown dwarf
binaries are extremely rare \citep{farihietal05-1}, so finding one
with a CPM WD companion appears rather unlikely, however infrared data
could rule this out for certain. The second option, a merger, can also
not be excluded. The mass of SDSS\,J1300+5904 is slightly below the
mean mass for single WDs, so any merger event would have
either been the merger of two low mass WDs, possibly helium core WDs,
or had to involve a low mass star. In the case that SDSS\,J1300+5904
is the product of a merger, one would expect the WD to be rapidly
spinning, which might be detected via photometric variability
\citep[e.g.][]{brinkworthetal04-1, brinkworthetal05-1}. Thus
SDSS\,J1300+5904 warrants further study.

\section{Conclusions}
\label{s-con}

We have shown how wide, non-interacting WD pairs can be used
to constrain the IFMR.  We have also shown that SDSS\,J1300+5904, the
CPM companion to the DA WD PG\,1258+593, is a MWD with
$B\simeq6$\,MG. The masses of both WDs are $\sim0.54\,\Msun$, slightly
below the average of non-magnetic WDs. Nevertheless, the two WDs
exhibit a significant difference in their effective temperatures,
implying an age difference of $\sim1.6$\,Gyr. Adopting standard
stellar evolution models, we show that assuming a progenitor mass of
$\sim1.5$\,\Msun\ for PG\,1258+593 implies a progenitor mass of
$\sim2-3$\,\Msun\ for SDSS\,J1300+5904, consistent with the mass of Ap
stars. An origin of the magnetic field related to common envelope
evolution is not impossible, but requires the assumption of an initial
triple system.

\section*{Acknowledglements}
\label{s-ack}

Work on magnetic WDs in Heidelberg was supported by grand 50\,OR\,0802
of the ``Deutsche Agentur fuer Luft und Raumfahrt''.

\bibliographystyle{mn_new}
\bibliography{aamnem99,aabib,proceedings,submitted,jon}

\bsp

\label{lastpage}

\end{document}